\documentclass[a4paper]{jpconf}

\usepackage{epsfig}
\def\cf4      {CF${_4}$\ }
\def\headtail      {``head-tail''\ }
\def\Cf      {$^{252}$Cf\ }

\usepackage{lineno}

\begin{document}

\title{
Improved measurement of the \headtail effect in nuclear recoils
}

\author{
D.~Dujmic$^3$,
S.~Ahlen$^1$,
P.~Fisher$^3$,
S.~Henderson$^3,$
A.~Kaboth$^3$,
G.~Kohse$^3$,
R.~Lanza$^3$,
M.~Lewandowska$^1$,
J.~Monroe$^3$,
A.~Roccaro$^1$,
G.~Sciolla$^3$,
N.~Skvorodnev$^2$,
H.~Tomita$^1$,
R.~Vanderspek$^3$,
H.~Wellenstein$^2$, 
R.~Yamamoto$^3$
}

\address{$^1$Boston University, Boston, MA 02215}
\address{$^2$Brandeis University,  Waltham, MA 02454}
\address{$^3$Massachusetts Institute of Technology, Cambridge, MA 02139}

\begin{abstract}
We present new results with a prototype detector that is being developed by the DMTPC collaboration for
the measurement of the direction tag (\headtail) of dark matter wind.
We use neutrons from a \Cf source to create low-momentum
nuclear recoils in elastic scattering with the residual gas nuclei.
The recoil track is imaged in low-pressure time-projection chamber with optical readout.
We measure the ionization rate along the recoil trajectory, which allows us to 
determine the direction tag of the incoming neutrons. 
\end{abstract}

%
%

\section{Introduction}
\label{sec::introduction}
The non-baryonic dark matter in the form of weakly interacting massive particles (WIMPs)
still eludes detection despite recent achievements in the detection technology~\cite{DMreview}. 
Aside from scaling up the size of existing detectors, the improvement in the detection sensitivity is 
possible by  detecting  the direction of the incoming dark matter particles. 
As the Earth moves in the galactic halo, the dark matter particles appear to come from Cygnus constellation.
The direction tag of the of the incoming particle, often referred to as the 
\headtail effect, increases the sensitivity of a directional detector by 
one order of magnitude~\cite{agreen}. 

In this paper we present improved results for tagging the direction 
of low-energy nuclear recoils created by neutrons from a $^{252}$Cf source
by using a time-projection chamber with optical readout. The neutrons
are used in lieu of the dark matter particles because they create similar distributions of recoil
energies and angles.
The measurement of directionality  tag relies on the fact that the 
ionization rate of  recoiling nuclei depends on their residual energy, 
and therefore the direction of the recoil can be tagged from the light distribution along the track.

\section{Experiment and Results}
\label{sec::results}

%
%

The detector is in more details described in~\cite{Dujmic:2007bd}.
The chamber utilizes   $10 \times 10~{\rm cm}^2$ wire frames. 
The  drift region between the cathode mesh and the ground 
wire plane is 2.6~cm with average electric field of 580~V/cm, while the 
amplification region between the  ground  and the anode wire plane (+2.2~kV) is about 3~mm. 
The pitch of the wires for the ground (anode) plane is 2~mm (5~mm) and 
the wire diameter is 15~$\mu$m (50~$\mu$m).
The chamber is filled with \cf4 at 200~Torr. 
The scintillation light is recorded with a cooled CCD camera 
equipped with a photographic lens that images approximately 2~cm$^2$ of the anode plane. 
The spread of pixel yields due to ADC noise and dark current is 
25 counts. Images are corrected for ADC bias and hot channels are identified and
excluded from analysis.

%
%
%
%
%

Neutrons are created in the fission of the \Cf nucleus, which occurs in approximately 3\% 
of all decays and produces 3.8 neutrons per fission~\cite{endf}. 
The radioactivity of our \Cf source is 3.4~mCi and
we estimate the total flux of $10^4$ neutrons per second into the solid angle ($10^{-3}$~sr) of the detector.
The wires of the tracking chamber are aligned with the direction of the neutron beam.
The recoil length projected to the wire axis is longer in case of WIMP scattering, therefore,
of \headtail effect in neutron scattering is expected to be harder. 
We take sequential 1-second exposures with the CCD camera.
We  reject images that have segments shorter than 0.7~mm, and recoil tracks that fall 
close to the boundary of the CCD field of view. 
The energy of the recoil segment is determined from the projection of the light intensity 
to the axis perpendicular to the wire. The relation between the light intensity and the energy
is determined using alpha particles that travel perpendicular to the wire
and deposit a known amount of energy.
The range of the recoil segment is calibrated using the known pitch of anode wires
and the observed distance between wires in the CCD images.

%
%
\begin{figure}[t]
\includegraphics[width=7.5cm]{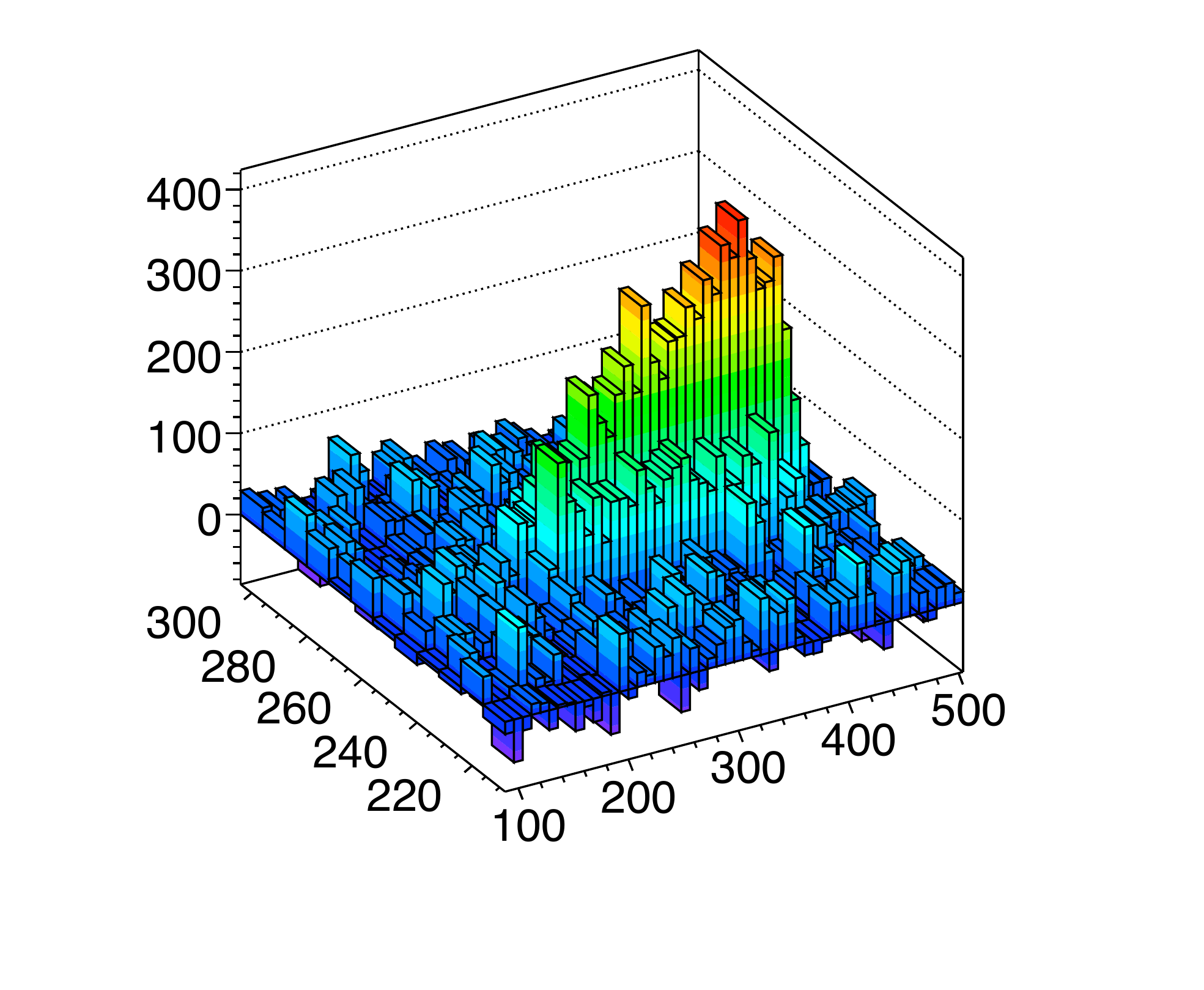} \hspace{2pc}%
\begin{minipage}[b]{7.5cm}
\caption{
An image of recoil track with neutrons coming from the right.
The noticeable asymmetry of the light yield along the wire 
indicates  of the \headtail effect.  \label{fg::recoil_images}
}
\end{minipage}
\end{figure}

An image of a nuclear recoil  in Figure~\ref{fg::recoil_images} shows 
noticeable asymmetry of the light yield along the wire.
In order to quantify this effect,  we define the skewness $\gamma$ 
as the dimensionless ratio between the third and second moments of the light yield along the wire coordinate ($x$): 
$
        \gamma = \frac{ \left< (x-\left<x\right>)^3 \right> }{ \left< (x-\left<x\right>)^2 \right>^{3/2}}. 
$
The sign indicates the slope of the light intensity along the track:
recoils that travel in the direction of the incoming neutrons have a negative skewness.

\begin{figure}[ht]
\center
\begin{tabular}{ll}
\includegraphics[width=7.5cm]{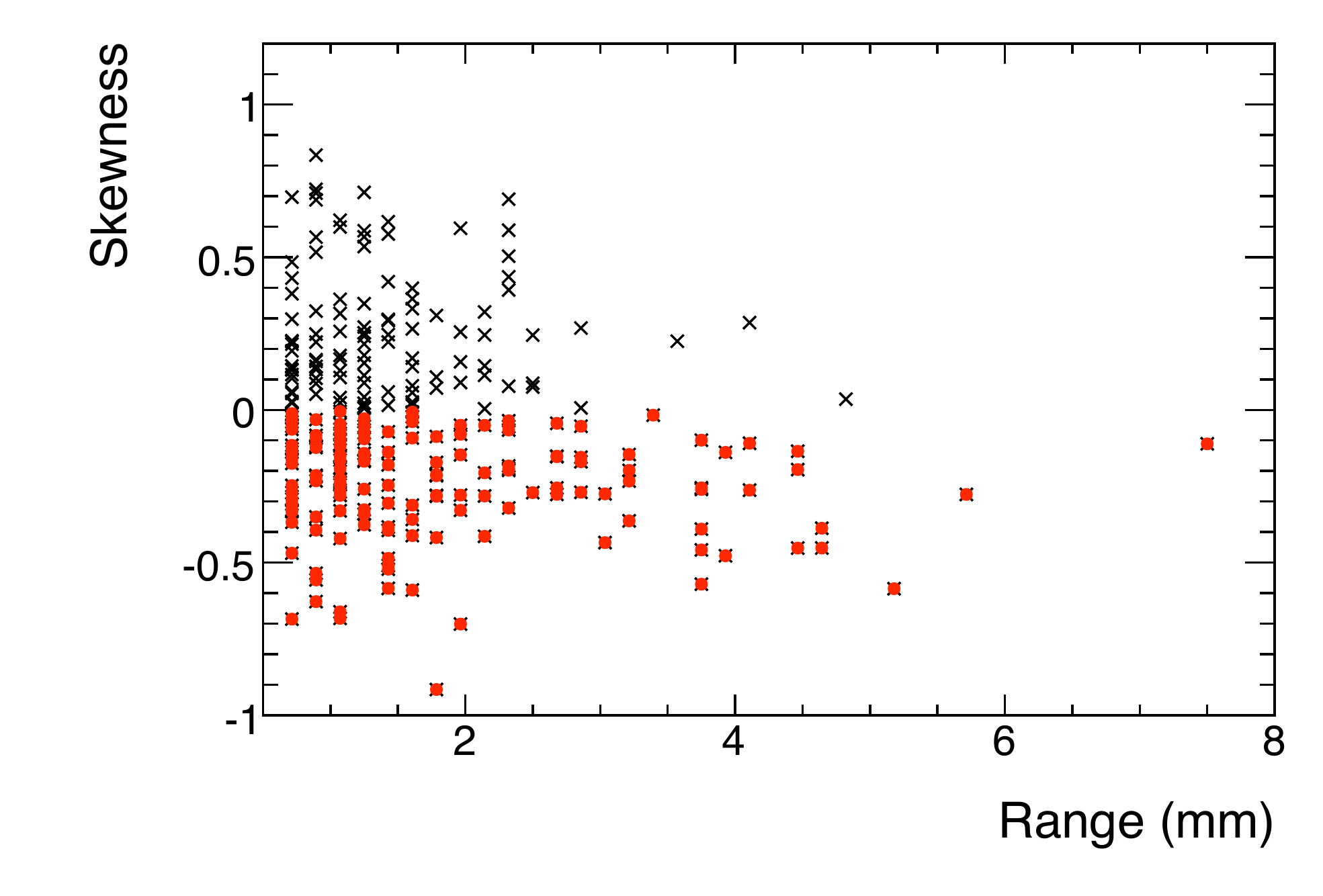} &
\includegraphics[width=7.5cm]{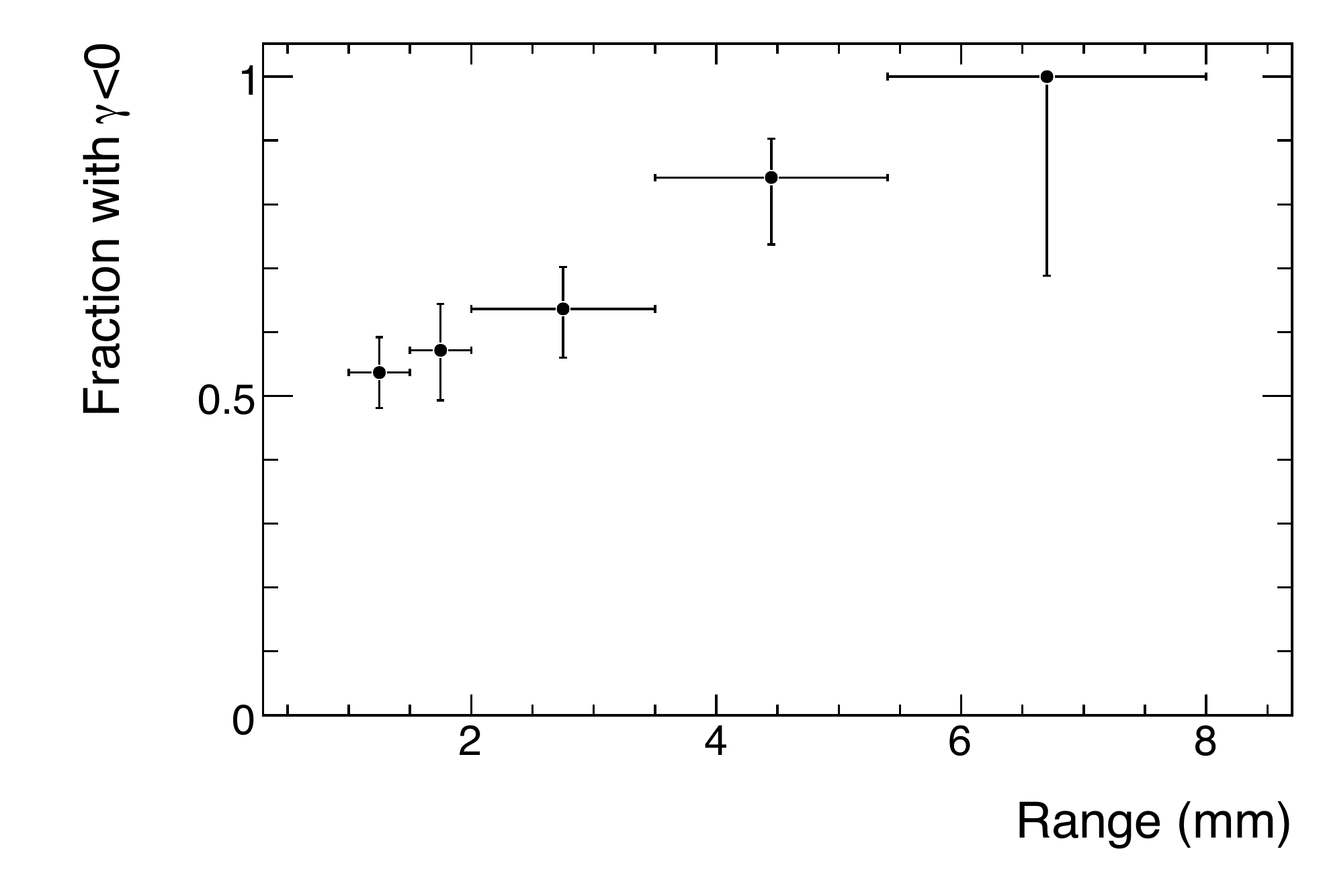}
\end{tabular}
\caption{Left: skewness as a function of the track length of the recoil segments.
Right: fraction of events with negative skewness as a function of the track length.
\label{fg::recoil_energy_vs_skewness}}
\end{figure}

A plot of the measured skewness as a function of the segment length is shown 
in the top plot of Figure~\ref{fg::recoil_energy_vs_skewness}.
The data in this plot corresponds to 3.6~h of live time using 5~mg of \cf4 gas.
The head-tail asymmetry is easier to observe for longer tracks that are better 
aligned with the anode wires and create more scintillation light. 
The bottom plot in Figure~\ref{fg::recoil_energy_vs_skewness}  shows 
the fraction of events with negative skewness as a function of the track length.

%
%

\section{Discussion and Conclusion }
\label{sec::summary}

\begin{figure}[hb]
\includegraphics[width=7.5cm]{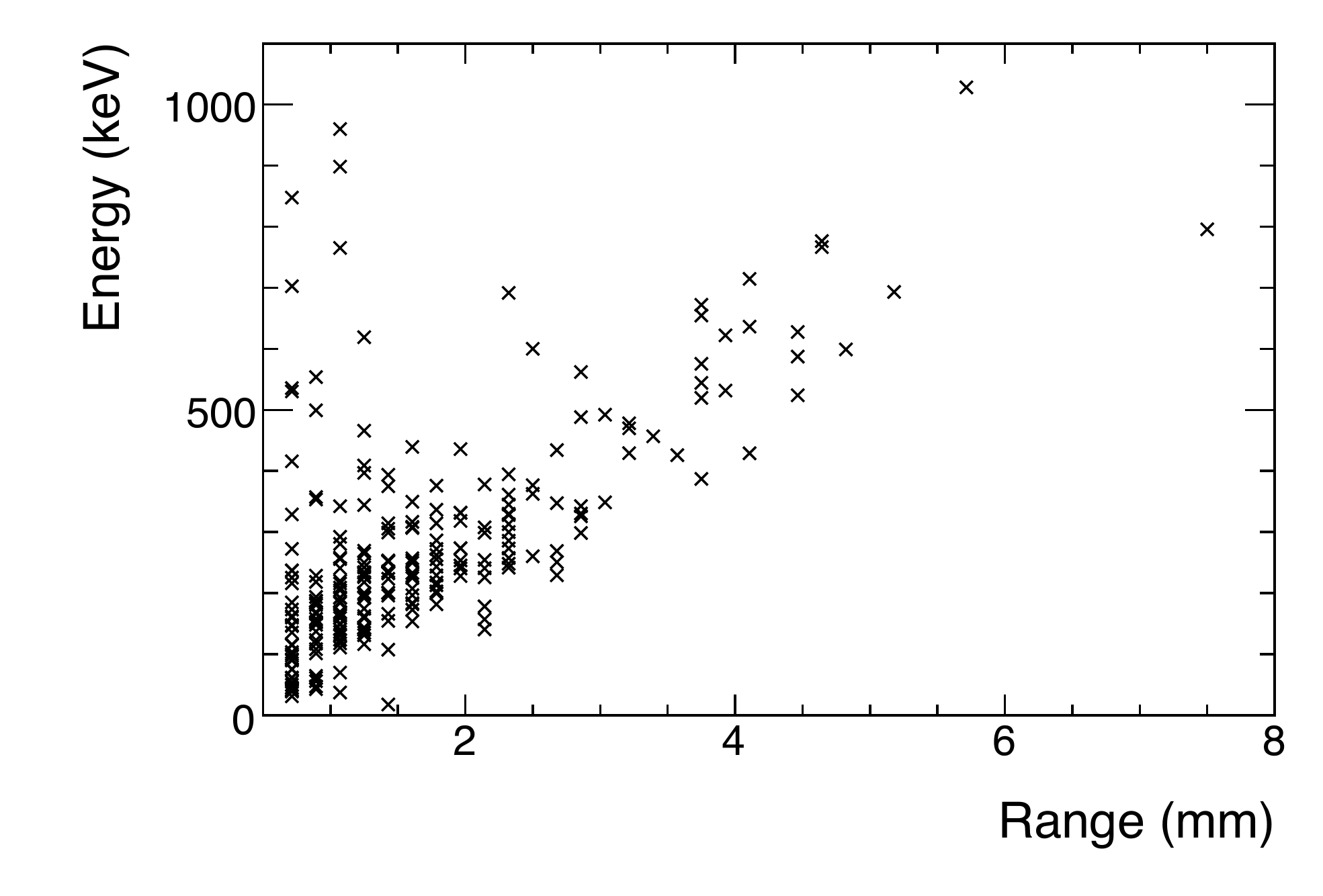} 
\hspace{2pc}%
\begin{minipage}[b]{7.5cm}\caption{
Correlation between the energy and the projected range of the recoil segments in data.
\label{fg::recoil_energy_vs_length}
}
\end{minipage}
\end{figure}

%
%
%
Since the measured light yield is proportional to the energy of the recoil segment and the length is proportional to
the track range projected to the wire, these two quantities should be correlated.
Figure~\ref{fg::recoil_energy_vs_length} shows  clear correlation between the 
light yield  versus length of the recoil segments.

We collect 1~day of live-time of data without sources and find two events that pass
our standard selection cuts. 
We verify good rejection of gammas by collecting 1/3~day of live-time of data with
$^{137}$Cs source (8~$\mu$Ci) placed near the sensitive area of our detector and
find zero events passing the cuts.

We assign a  conservative error of 10\% to the density of the \cf4 gas.
The statistical uncertainty on the energy measurements is about 10\%. 
The systematic error on the energy comes from non-uniformity in wire gain, stability of the gain over time, 
the pressure measurement and the calibration method that assumes the energy-independent proportionality of
the stopping power with the ionization rate.
The error on the recoil range comes from the analysis technique that overestimates
the range for low-energy recoils with the range close to the diffusion width.

%
%


We have presented improved results for tagging the direction of low-momentum nuclear 
recoils generated in the elastic scattering of low-energy neutrons with \cf4 gas. 
We have shown that in our current experimental setup the tag of incoming particle can 
be determined for recoil energies above 200~keV.
This threshold can be further reduced with expected improvements in the detector preformance.
This study has profound implications for the development of  dark matter detectors,
as the directionalty will be essential to produce convincing evidence 
for dark matter particles in the presence of backgrounds.

\end{document}